\documentclass{PoS}
\usepackage{amsfonts,amssymb,amsmath,bm}
\usepackage{graphicx}

\newcommand{\be}{\begin{equation}}
\newcommand{\bea}{\begin{eqnarray}}
\newcommand{\ee}{\end{equation}}
\newcommand{\eea}{\end{eqnarray}}
\newcommand{\bpi}{\begin{picture}}
\newcommand{\bce}{\begin{center}}
\newcommand{\epi}{\end{picture}}
\newcommand{\ece}{\end{center}}

\def\s#1{{\scriptscriptstyle #1}}

\def\gtreeb{\widetilde{\Gamma}^{(0)}}
\def\gfullb{\widetilde{\Gamma}}
\def\diff{{\rm d}}

\title{Impact of ghost loops on \\ dynamical gluon mass generation}

\ShortTitle{Impact of ghost loops on dynamical gluon mass generation}

\author{\speaker{Arlene~C.~Aguilar}\\
        Federal University of ABC, CCNH,\\ 
Rua Santa Ad\'elia 166,  CEP 09210-170, Santo Andr\'e, Brazil \\
        E-mail: \email{arlene.aguilar@ufabc.edu.br}}

\abstract{
Exploiting the gauge-invariant properties of the PT-BFM truncation scheme 
for the gluon Schwinger-Dyson equation,
we estimate the  individual non-perturbative contribution 
of the ``one-loop dressed'' ghost loop to the gluon 
propagator. Using the available quenched lattice 
data for the gluon and the ghost propagators  in  $d=4$ and $d=3$, 
we determine how the overall shape of the gluon propagator is affected by 
the removal of the ghost loop, and what are the consequences on the 
phenomenon of gluon mass generation.}

\FullConference{International Workshop on QCD Green's Functions, Confinement and Phenomenology,\\
		September 05-09, 2011\\
		Trento Italy}

\begin{document}

\section{Introduction}

In the last few years our understanding of the
infrared (IR) behavior of the fundamental QCD Green's function 
has improved substantially. Putting together  
the information obtained though various non-perturbative methods,  
such as lattice simulations~\cite{Cucchieri:2007md,
Cucchieri:2003di,Bogolubsky:2007ud,Oliveira:2008uf,arXiv:0912.0437,Bowman:2007du},
Schwinger-Dyson  equations (SDEs)~\cite{Aguilar:2008xm, Binosi:2009qm, Boucaud:2008ji}, 
functional methods~\cite{Braun:2007bx,Szczepaniak:2010fe},
and algebraic  techniques~\cite{Dudal:2008sp,Kondo:2011ab}, 
it is  by now well-established that, in the Landau gauge, the  gluon propagator  and a ghost dressing function 
are finite in the IR 
(in d = 3,4).~\cite{Aguilar:2008xm,Boucaud:2008ji}. 
Evidently, these results support the gluon mass generation picture 
proposed by Cornwall several years ago~\cite{Cornwall:1981zr},  
disfavoring the so-called ``ghost-dominance'' 
picture of QCD~\cite{Alkofer:2000wg,Zwanziger:2001kw}, 
whose theoretical cornerstone has been the 
existence of a divergent (``IR-enhanced'') ghost dressing function.

However, the finiteness of the ghost sector does not imply necessarily that the ghost 
contribution has been relegated to a marginal role in the QCD dynamics. 
In fact, compelling evidence to the contrary 
has emerged from  
detailed studies of the gap equation that controls
the breaking of chiral symmetry and the dynamical generation 
of a constituent quark mass~\cite{Fischer:2003rp,Aguilar:2010cn}. Specifically, 
a detailed study has revealed 
that the proper inclusion of the ghost sector into the quark SDE
is  crucial for obtaining quark masses of the order of $300$ MeV in the presence
of finite gluon propagator ~\cite{Aguilar:2010cn}

Given the importance of the ghost sector for the dynamical generation 
of a constituent quark mass, the main purpose of this talk 
is to ask whether a similar situation applies  
in the case of the dynamical generation of an effective gluon mass~\cite{Aguilar:2011yb}.

\section{Disentangling the ``one-loop dressed'' ghost contributions}

In what follows we will work within the 
specific framework provided by the  synthesis of the pinch technique (PT)
~\cite{Cornwall:1981zr,Cornwall:1989gv,Binosi:2009qm,Aguilar:2006gr,Binosi:2007pi} 
with the background field method (BFM)~\cite{Abbott:1980hw}. 

\begin{figure}[!t]
\begin{center}
\includegraphics[scale=1.0]{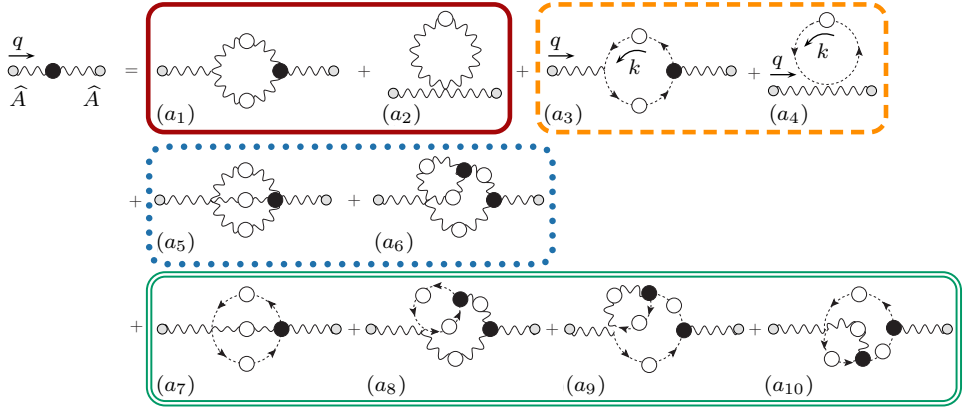}
\caption{\label{fig1}The SDE corresponding to the PT-BFM gluon self-energy
$\Pi^{ab}_{\mu\nu}(q)$. The graphs inside each box 
furnish an individually transverse contribution. White
(black) circles denote full propagators (vertices).}
\end{center}
\end{figure}

We start by recalling that within the PT-BFM scheme the  gluon self-energy, $\Pi_{\mu\nu}(q)$, 
is giving by the sum of the diagrams represented by Fig.~\ref{fig1}, i.e. 
\be
\Pi^{\mu\nu}(q) = \sum_{i=1}^{10} (a_i)^{\mu\nu}\,.
\ee

Exploiting the well-known blockwise transversality properties of the PT-BFM $\Pi^{\mu\nu}(q)$, 
it is possible to separate the transversal 
contribution of the one-loop dressed ghost diagrams, represented by the
diagrams $(a_3)$ and $(a_4)$, and to be denoted by
\be
\Pi_c^{\mu\nu}(q)=(a_3)^{\mu\nu}+(a_4)^{\mu\nu} \,;
\label{Pica}
\ee
where $(a_3)$ and $(a_4)$ are given by 
\bea
(a_3)_{\mu\nu}&=& -g^2C_A
\int_k\!\gtreeb_{\mu}(k,q,-k-q)D(k)D(k+q)\gfullb_{\nu}(k+q,-q,-k), 
\nonumber \\
(a_4)_{\mu\nu}&=&2 g^2C_A g_{\mu\nu}\int_k\!D(k).
\label{gh-1ldr}
\eea
In the equations above, $D^{ab}(q^2)=\delta^{ab}D(q^2)$ denotes the full ghost propagator, 
defined in terms of the ghost dressing function $F(q^2)$ as
\be
D(q^2)= \frac{F(q^2)}{q^2},
\label{ghdr}
\ee
while $\gfullb_{\mu}$ represents the three-particle vertex describing 
the interaction of the background gluon with a ghost and an antighost, 
with (all momenta entering)
\be
i\Gamma_{c^b \widehat{A}^a_\mu \bar c^c}(r,q,p)=gf^{acb}\widetilde{\Gamma}_{\mu}(r,q,p); \qquad \widetilde{\Gamma}^{(0)}_{\mu}(r,q,p)=(r-p)_\mu.
\ee
Finally, $C_A$ is the Casimir eigenvalue of the adjoint representation	
[$C_A=N$ for $SU(N)$], and we have introduced the $d$-dimensional 
integral measure (in dimensional regularization) according to
\be
\int_{k}\equiv\frac{\mu^{\epsilon}}{(2\pi)^{d}}\!\int\!\mathrm{d}^d k,
\label{dqd}
\ee
with $\mu$ the 't Hooft mass, and $\epsilon=4-d$.
Then, by virtue of the PT-BFM Ward identity 
\be
iq^\mu\gfullb_\mu(r,q,p)=D^{-1}(r)-D^{-1}(p),
\label{WI}
\ee
it is immediate to establish the transversality of $\Pi_c^{\mu\nu}(q)$, namely~\cite{Aguilar:2006gr}
\be
q_{\mu}\Pi_c^{\mu\nu}(q) =0.
\ee

%%%%%%%%%%%%%%%%%%%%%%%%
%  Fig1 
%%%%%%%%%%%%%%%%%%%%%%%%%%%%%%%%%%%%%%%%%%%%%%%%%%%%%%%
%%%%%%%%%%%%%%%%%%%%%%%%%%%%%%%%%%%%%%%

%
It is convenient for our purposes to decompose the full self-energy $\Pi_{\mu\nu}(q)$ as 
\be
\Pi^{\mu\nu}(q) = \Pi^{\mu\nu}_r(q) + \Pi_c^{\mu\nu}(q) \,,
\label{sff}
\ee
where  $\Pi^{\mu\nu}_r(q)$ denotes the sum 
of the remaining subsets of diagrams in  
Fig.~\ref{fig1}, {\it i.e.}, both the gluon one- and two-loop dressed diagrams, 
as well as two-loop dressed ghost diagrams, 
\be
\Pi^{\mu\nu}_r(q) = \sum_{\substack{i=1\\i\neq3,4}}^{10} (a_i)^{\mu\nu}\,.
\ee
Notice that  due to the special Ward identities satisfied by the PT-BFM vertices, 
$\Pi^{\mu\nu}_r(q)$ is also transverse~\cite{Aguilar:2006gr,Binosi:2007pi}.

Using Eq.~(\ref{sff}), the SDE for the full gluon propagator in the Landau gauge of the PT-BFM scheme 
assumes then the form~\cite{Aguilar:2008xm} 
\be
\Delta^{-1}(q^2)P^{\mu\nu}(q)=\frac{q^2P^{\mu\nu}(q)+i \left[\Pi^{\mu\nu}_r(q)+\Pi^{\mu\nu}_c(q)\right]}{\left[1+G(q^2)\right]^2},
\label{gSDE}
\ee 
where the gluon propagator $\Delta_{\mu\nu}(q)$ is defined as 

\be
\Delta_{\mu\nu}(q)=-i\Delta(q^2)P_{\mu\nu}(q); \qquad
P_{\mu\nu}(q)=g_{\mu\nu}-\frac{q_\mu q_\nu}{q^2},
\ee
The function $G$ appearing in (\ref{gSDE}) is the form factor associated with $g_{\mu\nu}$
in the Lorentz decomposition of the auxiliary two-point function $\Lambda$, given by~\cite{Aguilar:2008xm,Aguilar:2009nf} 
\bea
\Lambda_{\mu\nu}(q)&=&-ig^2C_A\int_k\!\Delta_\mu^\sigma(k)D(q-k)H_{\nu\sigma}(-q,q-k,k)\nonumber\\
&=&g_{\mu\nu}G(q^2)+\frac{q_\mu q_\nu}{q^2}L(q^2).
\eea

Notice that the auxiliary function $H$ is  related to the (conventional) gluon-ghost vertex by the identity
\be
\Gamma_{\mu}(r,q,p)=-p^\nu H_{\nu\mu}(p,r,q),
\ee
and that, in the (background) Landau gauge,  
 the following all order relation holds~\cite{Grassi:2004yq,Aguilar:2009pp}
\be
F^{-1}(q^2)=1+G(q^2)+L(q^2).
\label{funrel}
\ee 

Now, let us return to Eq.~(\ref{gSDE}), 
and define in a  completely analogous way the quantity $\Delta_r(q^2)$, given by 
\be
\Delta_r^{-1}(q^2)P^{\mu\nu}(q)=\frac{q^2P^{\mu\nu}(q)+i\Pi^{\mu\nu}_r(q)}{\left[1+G(q^2)\right]^2}. 
\label{Dr}
\ee
Evidently, $\Delta_r$ represents the propagator 
obtained by subtracting out from the full propagator 
$\Delta$ the one-loop dressed ghost contributions. 
Then, taking the trace of both Eqs.~(\ref{gSDE}) and (\ref{Dr}), 
defining the trace of $\Pi^{\mu\nu}_c(q)$ as 
\be
\Pi_c(q^2) \equiv \Pi^{\mu}_{c\,\mu}(q) , 
\label{Pic}
\ee
 and solving for $\Delta_r$, we arrive at 
\be
\Delta_r(q^2)=\Delta(q^2)\left\{1-\frac{i\Delta(q^2)\Pi_c(q^2)}{(d-1)\left[1+G(q^2)\right]^2}\right\}^{-1},
\label{me}
\ee
which represents our master formula.

In order to obtain the behavior of the  
propagator $\Delta_r(q^2)$ from Eq.~(\ref{me}) we will in 
the next sections 
{\it (i)} identify the full gluon propagator $\Delta(q^2)$
with that obtained from the lattice, and {\it (ii)} 
determine nonperturbatively the quantity $\Pi_c$ 
from Eqs.~(\ref{gh-1ldr}) and (\ref{Pic}), and evaluate it numerically 
using as input the lattice results 
for the ghost dressing function $F(q^2)$.

\section{ The nonperturbative calculation of $\Pi_c(q^2)$}   

The first step in the calculation of the non-perturbative quantity $\Pi_c(q^2)$  
is the definition of an  Ansatz for the fully-dressed ghost vertex $\gfullb_{\mu}$, appearing 
in graph $(a_3)$ of Eq.~(\ref{gh-1ldr}) which satisfies 
the crucial Ward identity of Eq.~(\ref{WI}). 
This task can be accomplished with the help of the ``gauge-technique''~\cite{Salam:1963sa}
which reconstruct the vertex by ``solving'' 
its Ward identity. Following the same steps of the derivation presented in~\cite{Ball:1980ay} for
the scalar QED vertex, the  Ansatz for the fully-dressed ghost vertex $\gfullb_{\mu}$,
reads,
\be
\gfullb_\mu(r,q,p)=i\frac{(r-p)_\mu}{r^2-p^2}\left[D^{-1}(p^2)-D^{-1}(r^2)\right],
\label{va}
\ee 
which evidently satisfies Eq.~(\ref{WI}) when contracted with $q^{\mu}$. Obviously the  ``gauge technique'' 
leaves the transverse (automatically conserved) part of the vertex undetermined, which, on general grounds,  
is expected to be subleading in the IR~\cite{Salam:1963sa, Kizilersu:2009kg}. 

Substituting~(\ref{va}) in the first equation of~(\ref{gh-1ldr}) and taking the trace,
it is relatively straightforward to obtain the result
\be
\Pi_c(q^2) = g^2C_A \left[4T(q) - q^2R(q)\right],
\label{PicTR}
\ee
where 
\bea
R(q)&=&\int_k\!\frac{D(k+q)-D(k)}{(k+q)^2-k^2},\nonumber \\
T(q)&=&\int_k\!k^2\frac{D(k+q)-D(k)}{(k+q)^2-k^2}+\frac d2\int_k\!D(k).
\label{RandT}
\eea
To further evaluate $\Pi_c(q^2)$, we must invoke 
the so-called ``seagull-identity''~\cite{Aguilar:2009ke},  
\be
\int_k\! k^2\frac{\partial{f}(k^2)}{\partial k^2}+\frac d2\int_k\!f(k^2)=0, 
\label{seagull}
\ee 
valid in dimensional regularization, which enforces the cancellations of all seagull-type of divergences.
Notice that,  in the limit  $q\to 0$, the term $q^2R(q)$ vanishes, and so does $T(q)$, since 
\bea
T(q)\ \stackrel{q\to0}{\to}\ T(0) &=&
\int_k\!k^2\frac{\partial D(k^2)}{\partial k^2}+\frac d2\int_k\!D(k) = 0 , 
\label{T0}
\eea
where in the last step we have employed Eq.~(\ref{seagull}), with \mbox{$f(k^2) \to D(k^2)$}. 
Employing this result, follows immediately from Eq.~(\ref{PicTR}) that $\Pi_c(0) =0$. Using
the fact that \mbox{$\Delta^{-1}(0) = m^2(0)$}, we conclude that   the one-loop
dressed ghost diagrams $(a_3)$ and $(a_4)$ do not contribute {\it directly} to the value of 
dynamical gluon mass at zero momentum transfer.  
The easiest way to appreciate this is by recalling that the mechanism responsible for 
endowing the gluon with a dynamical mass relies on the presence of massless poles in the 
nonperturbative tree-gluon
[the black circle in graph $(a_1)$ of Fig.~\ref{fig1}], whereas the ghost vertex has the usual structure
[note the absence of poles in the Ansatz of Eq.~(\ref{va})]~\cite{Aguilar:2011ux}. 

In addition, notice that  when $d=4$,  $R(q)$ is 
ultraviolet divergent, and must be properly renormalized, by introducing the appropriate wave-function renormalization
constant. 

The  (subtractive) renormalization must be carried out at the level of (\ref{gSDE}).
Specifically (setting directly $d=4$),  
\be
\Delta^{-1}(q^2)=\frac{Z_A q^2+\frac{i}{3}\left[\Pi_r(q) +\Pi_c(q)\right]}{\left[1+G(q^2)\right]^2},
\label{rgSDE}
\ee 
where
the renormalization constant $Z_A$ is fixed in the MOM scheme through the
condition \mbox{$\Delta^{-1}(\mu^2) = \mu^2$}. 
Applying this condition at the level of Eq.~(\ref{rgSDE}) together with Eq.~(\ref{funrel}), and
using the fact that the function   
$L(x)$ is considerably smaller  than $G(x)$ in the entire range of momenta, (so that we can use the approximation $1+G(\mu^2) \approx F^{-1}(\mu^2)=1$)  allows one to express $Z_A$ as
\be
Z_A= 1 - \frac{i}{3\mu^2}\left[\Pi_{r}(\mu) + \Pi_{c}(\mu) \right].
\label{z12}
\ee
Finally, substituting Eq.~(\ref{z12}) into Eq.~(\ref{rgSDE}), and defining (in a natural way)   
the renormalized $\Delta^{-1}_r(q^2)$ as 
\bea
\Delta^{-1}_r(q^2)= \frac{ q^2 + \frac{i}{3}\left[\Pi_{r}(q) - (q^2/\mu^2)\Pi_{r}(\mu)\right]}{[1+G(q^2)]^2},
\label{deltag_re}
\eea 
the renormalized version of the master formula (\ref{me}) will read
\be
\Delta_{r}^{-1}(q^2) = \Delta^{-1}(q^2) - 
\frac{i}{3}\frac{\left[\Pi_{c}(q) - (q^2/\mu^2)\Pi_{c}(\mu)\right]}{[1+G(q^2)]^2}.  
\label{rme}
\ee
Evidently (\ref{rme}) is obtained from (\ref{me}) by replacing  $\Delta^{-1}(q^2) \to \Delta^{-1}_{R}(q^2)$  
(``$R$'' for ``renormalized''), and  $\Pi_{c}(q) \to  \Pi_{c, R}(q)$, where  
\be
\Pi_{c, R}(q) = \Pi_{c}(q) - (q^2/\mu^2)\Pi_{c}(\mu).
\label{rpic}
\ee

For the ensuing numerical treatment of $R(q)$ and $T(q)$
carried out in the next section, it is advantageous to have the 
crucial property $T(0)=0$ a priori built in, 
in order to avoid possible deviations due to minor numerical instabilities. 
To that end, we introduce the quantity $\overline{T}$
\bea
\overline{T}(q)=T(q)-T(0) =\int_k\!k^2\left[\frac{D(k+q)-D(k)}{(k+q)^2-k^2}-\frac{\partial D(k)}{\partial k^2}\right],
\eea
which has the property of ensuring (by construction) that $\overline{T}(0)=0$, while, at the same time, 
coinciding with the original $T$ for all momenta $q$.

In addition, it is convenient to re-express $R(q)$ and $\overline{T}(q)$ 
in terms of the ghost dressing function. 
Using  Eq.~(\ref{ghdr}), 
after some elementary algebra, one obtains
\bea
R(q)=-\int_k\frac{F(k)}{k^2(k+q)^2}+\int_k\!\frac{F(k+q)-F(k)}{k^2[(k+q)^2-k^2]}, 
\,\,\,\,
\overline{T}(q)= 
\int_k \left[\frac{F(k+q)-F(k)}{(k+q)^2-k^2}-\frac{\partial F(k)}{\partial k^2}\right]\!;
\label{finalform}
\eea 
note that the angular integration of the first term in $R$ can be carried out 
analytically for any value of the space-time dimension $d$. 

Finally, note that up until this point we have been working in 
Minkowski space. To make the transition to Euclidean space, 
we must employ the usual rules. Specifically, 
we set $\int_k\!=\mathrm{i}\!\int_{k_\mathrm{\s E}}$ and $q^2_\mathrm{\s E} = -q^2$, 
and use that 
$\Delta_\mathrm{\s E}(q^2_\mathrm{\s E})=-\Delta(-q^2_\mathrm{\s E}); F_\mathrm{\s E}(q^2_\mathrm{\s E})=  F(-q^2_\mathrm{\s E});
G_\mathrm{\s E}(q^2_\mathrm{\s E})= G(-q^2_\mathrm{\s E}), $
suppressing the subscript ``E'' in what follows.

\section{Numerical Results}

We will now proceed to perform the numerical analysis. Using the available lattice data on the
ghost dressing function $F$,  we evaluate the terms $R$ and $\overline{T}$ given in 
Eq.~(\ref{finalform}),  
and combine them following the 
Eqs.~(\ref{PicTR}) and (\ref{rpic}) 
 to obtain the (renormalized) ghost contribution to the 
gluon self-energy $\Pi_c$.
Finally, we construct $\Delta_r$ using~(\ref{me}) and the
lattice results available for the gluon propagator $\Delta$. This exercise is carried out  
for two different cases: $d=4$, $N=3$,  and  $d=3$, $N=2$. 

%%%%%%%%%%%%%%%%%%%%%%%%%%%%%%%%%%%%%%%%%%%%%%%%%%%%%%%%%%%%%%%%%%%%%%%%%%
%             Fig.2  gluon propagator and ghost dressing SU(3)
%%%%%%%%%%%%%%%%%%%%%%%%%%%%%%%%%%%%%%%%%%%%%%%%%%%%%%%%%%%%%%%%%%%%%%%%%%%%
\begin{figure}[!t]
\begin{center}
\begin{minipage}[b]{0.45\linewidth}
\centering
\includegraphics[scale=0.5]{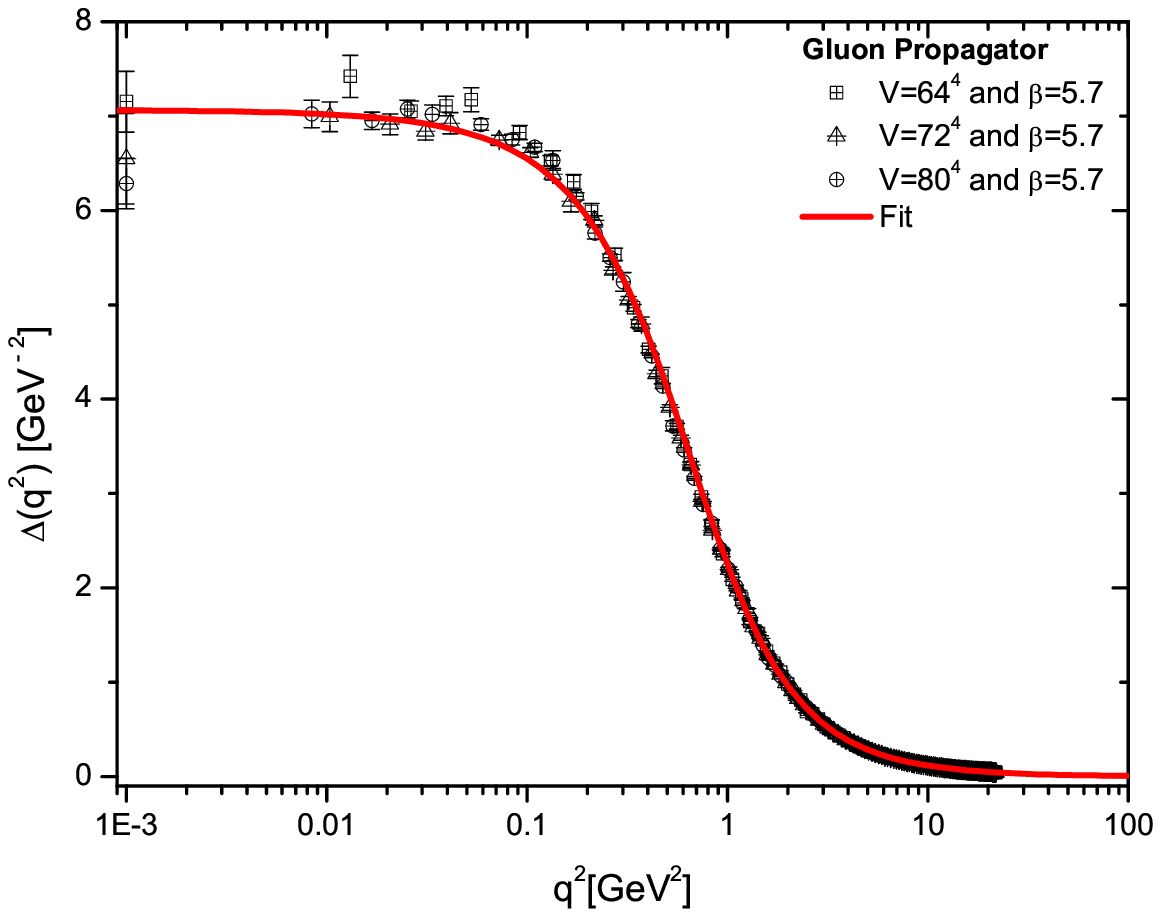}
\end{minipage}
\hspace{0.5cm}
\begin{minipage}[b]{0.50\linewidth}
%\hspace{-1.5cm}
\includegraphics[scale=0.5]{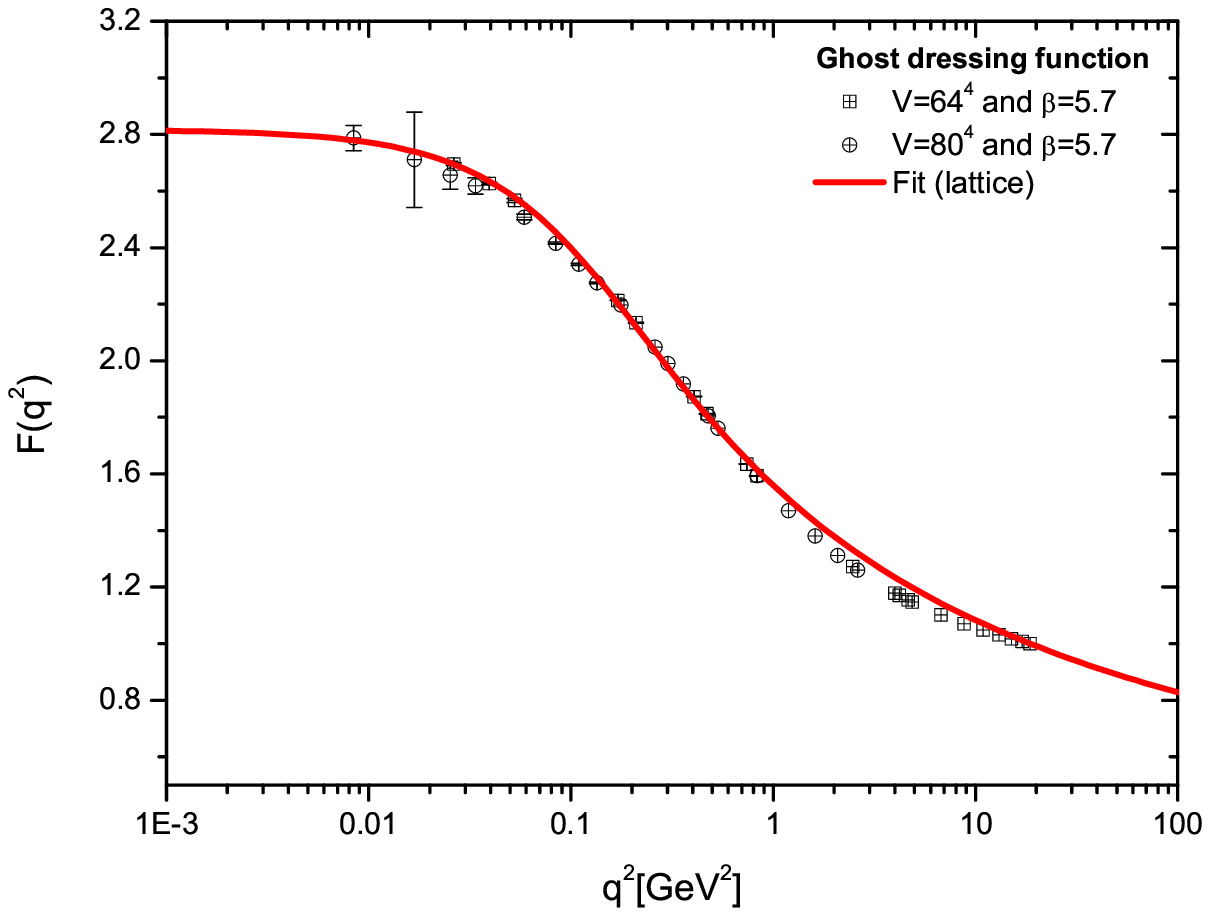}
\end{minipage}
\end{center}
\vspace{-0.5cm}
\caption{\label{ggh-4dSU3}{\it Left panel}: Lattice result for the $SU(3)$ gluon 
propagator, $\Delta(q)$, in $d=4$, renormalized  at \mbox{$\mu=4.3$ GeV}. The continuous 
line represents the fit given by Eq.~(4.1). {\it Right panel}: The $SU(3)$ ghost dressing
function, $F(q^2)$, renormalized at the same point, \mbox{$\mu=4.3$ GeV}; the solid line
corresponds to the fit given by Eq.~(4.2).}
\end{figure}
%%%%%%%%%%%%%%%%%%%%%%%%%%%%%%%%%%%%%%%%%%%%%%%%%%%%%%%%%%%%%%%%%%%%

In Fig.~\ref{ggh-4dSU3} we show the lattice results for the four-dimensional $SU(3)$ 
gluon propagator $\Delta(q^2)$ (left panel), and the  corresponding ghost dressing
function $F(q^2)$ (right panel), 
obtained from~\cite{Bogolubsky:2007ud}, and renormalized at 
$\mu=4.3$ GeV. 

As has been discussed in detail in the 
literature~\cite{Aguilar:2010cn,Aguilar:2011ux,Aguilar:2010gm}, both sets of data 
can be accurately fitted in terms of IR-finite quantities. 
More specifically, for the 
case of $\Delta(q^2)$, we have  proposed
a fit of the form~\cite{Aguilar:2010gm}
\be
\Delta^{-1}(q^2)= M^2(q^2) + q^2\left[1+ \frac{13C_{\rm A}g_1^2}{96\pi^2} 
\ln\left(\frac{q^2 +\rho_1\,M^2(q^2)}{\mu^2}\right)\right], \quad  M^2(q^2) = \frac{m_0^4}{q^2 + \rho_2 m_0^2}.
\label{gluon}
\ee

Notice that in the above expression, the finiteness of $\Delta^{-1}(q^2)$ is assured  
by the presence of the function $M^2(q^2)$, which forces the value 
of \mbox{$\Delta^{-1}(0) = M^2(0) = m_0^2/\rho_2$}.
The continuous line on the left panel of Fig.~\ref{ggh-4dSU3} 
corresponds our best fit, which can be 
reproduced setting \mbox{$m_0 = 520$~MeV}, $g_1^2=5.68$, $\rho_1=8.55$  and  $\rho_2=1.91$.

The $SU(3)$ lattice data for $F(q^2)$, 
shown in the right panel of Fig.~\ref{ggh-4dSU3}, will be fitted by 
the following expression 
\be
F^{-1}(q^2)= 1+ \frac{9}{4}\frac{C_{\rm A}g_1^2}{48\pi^2}
\ln\left(\frac{q^2 +\rho_3  M^2(q^2)}{\mu^2}\right);\qquad M^2(q^2) = \frac{m_0^4}{q^2 + \rho_4 m_0^2},
\label{ghdr-fit}
\ee
with the parameters given by \mbox{$m_0 = 520$~MeV}, $g_2^2 = 8.65$, $\rho_3 = 0.25$ and $\rho_4 = 0.64$. 
Notice that the $M(q^2)$ has the 
same power-law running as the one reported in Refs~\cite{Lavelle:1991ve,Aguilar:2007ie,Oliveira:2010xc}.

The only missing ingredient for the actual nonperturbative 
determination of $\Pi_c$, and therefore $\Delta_r$, is the value of \mbox{$\alpha_s=g^2/4\pi$}. 
Instead of choosing a single value for $\alpha_s$, we will
use the  physically motivated range of values [0.20,0.29], which will furnish a more representative 
picture of the numerical impact of the ghost corrections on the gluon 
propagator. 

\begin{figure}[!t]
\begin{center}
\begin{minipage}[b]{0.45\linewidth}
\centering
%\hspace{-1cm}
\includegraphics[scale=0.5]{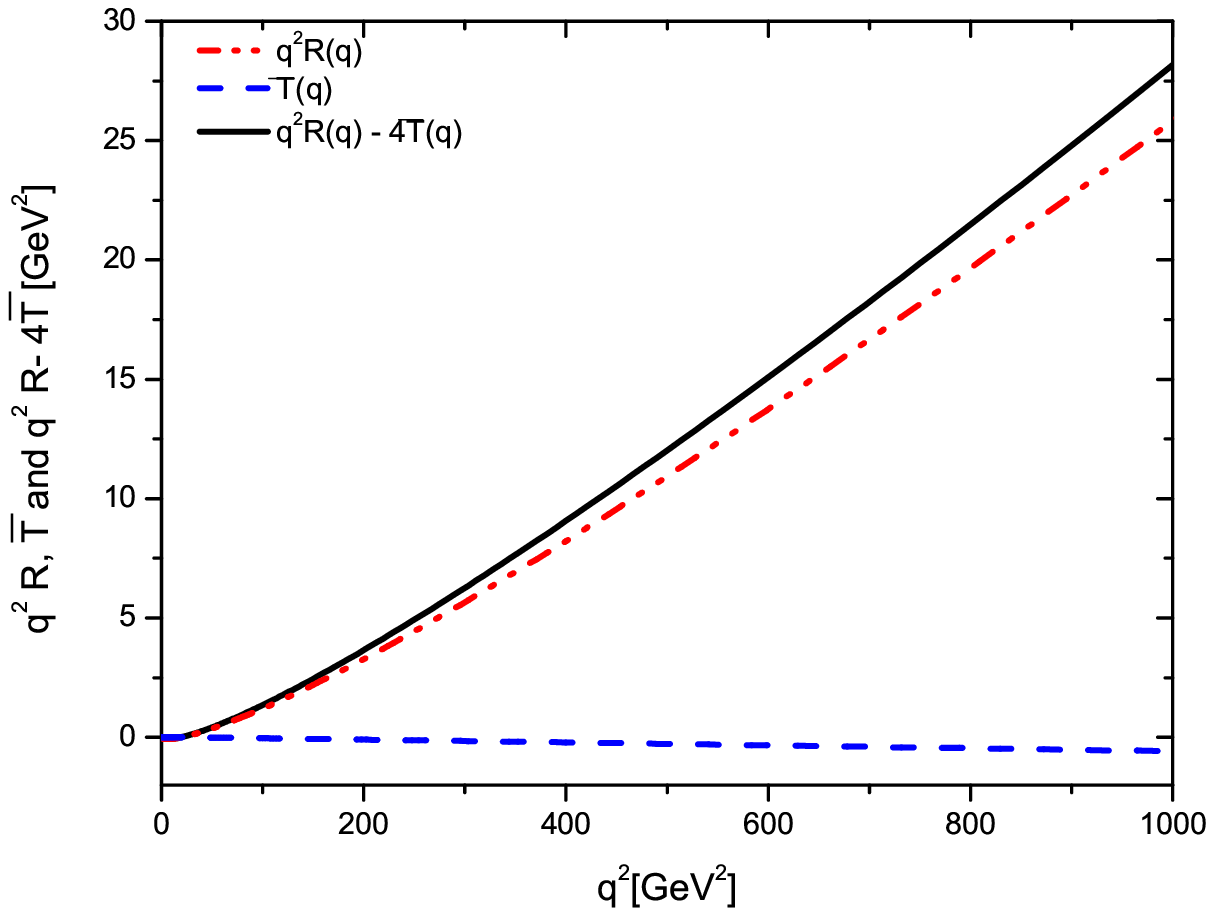}
\end{minipage}
\hspace{0.5cm}
\begin{minipage}[b]{0.50\linewidth}
%\hspace{-1.5cm}
\includegraphics[scale=0.5]{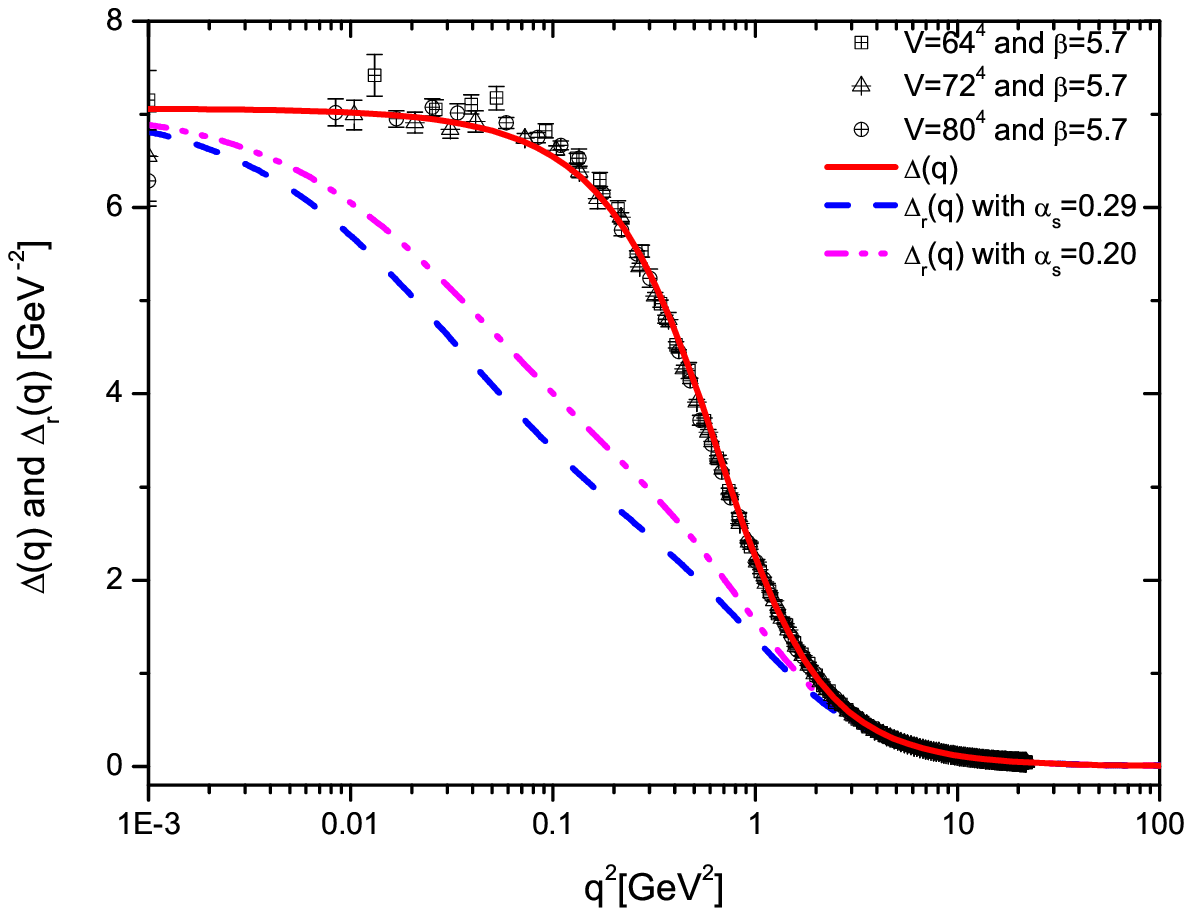}
\end{minipage}
\end{center}
\vspace{-0.5cm}
\caption{\label{RTandGluon-4dSU3}{\it Left panel}: Numerical evaluation 
of the ghost contribution $\Pi_c(q)$ to the gluon propagator using as input
 our best fit for the $d=4$, $N=3$ ghost dressing lattice data. {\it Right panel}: 
The removal of the one-loop dressed ghost contribution from the (lattice) 
gluon propagator  results in a diminished ``swelling'' in the momentum region below 1 GeV$^2$.} 
\label{GdressandChi-4dSU3}
\end{figure}

The results obtained for the renormalized $R$ and $\overline{T}$,  
after substituting into the corresponding formulas our best fit
for $F$, given by Eq.~(\ref{ghdr-fit}), are shown on the left panel of Fig.~\ref{RTandGluon-4dSU3}, together 
with the combination  $q^2 R -4 \overline{T}$, which appears on the rhs of 
Eq.~(\ref{PicTR}). It is clear that the contribution of the term $4 \overline{T}$ 
is rather negligible; in a way this is to be expected, given that this term 
vanishes identically in perturbation theory (for all values of $q$), and 
vanishes nonperturbatively at the origin.   

%%%%%%%%%%%%%%%%%%%%%%%%%%%%%%%%%%%%%%%%%%%%%%%%%%%%%%%%%%%%%%%%%%%%%%%%%%
%             Fig.3 \Pi_c and \Delta_r in 4d SU(3)
%%%%%%%%%%%%%%%%%%%%%%%%%%%%%%%%%%%%%%%%%%%%%%%%%%%%%%%%%%%%%%%%%%%%%%%%%%%%
%%%%%%%%%%%%%%%%%%%%%%%%%%%%%%%%%%%%%%%%%%%%%%%%%%%%%%%%%%%%%%%%%%%%
Next, we use these results to construct $\Pi_c$, given in Eq.~(\ref{rpic}), and 
finally $\Delta_r$, expressed by Eq.~(\ref{rme}) (Fig.~\ref{RTandGluon-4dSU3} right panel), using both 
values of $\alpha_s$, namely \mbox{$\alpha_s=0.29$} (blue dashed line) and 
\mbox{$\alpha_s=0.20$} (magenta dashed-dotted line). 

We then see that the net effect of removing the ghost contribution is to suppress 
significantly  
the support of the gluon propagator in the region below \mbox{1 GeV$^2$}. 
Higher values of $\alpha_s$  increase the impact of the ghost contributions, 
but only slightly,  as can be seen on the right panel of Fig.~\ref{RTandGluon-4dSU3}. 
As we will see in the next section, this ``deflating'' of the gluon propagator 
in the intermediate region of momenta, 
produced by the removal of the ghost contributions,  
has far-reaching consequences on the generation of a dynamical gluon mass. 

\begin{figure}[!t]
\begin{center}
\begin{minipage}[b]{0.45\linewidth}
\centering
%\hspace{-1cm}
\includegraphics[scale=0.5]{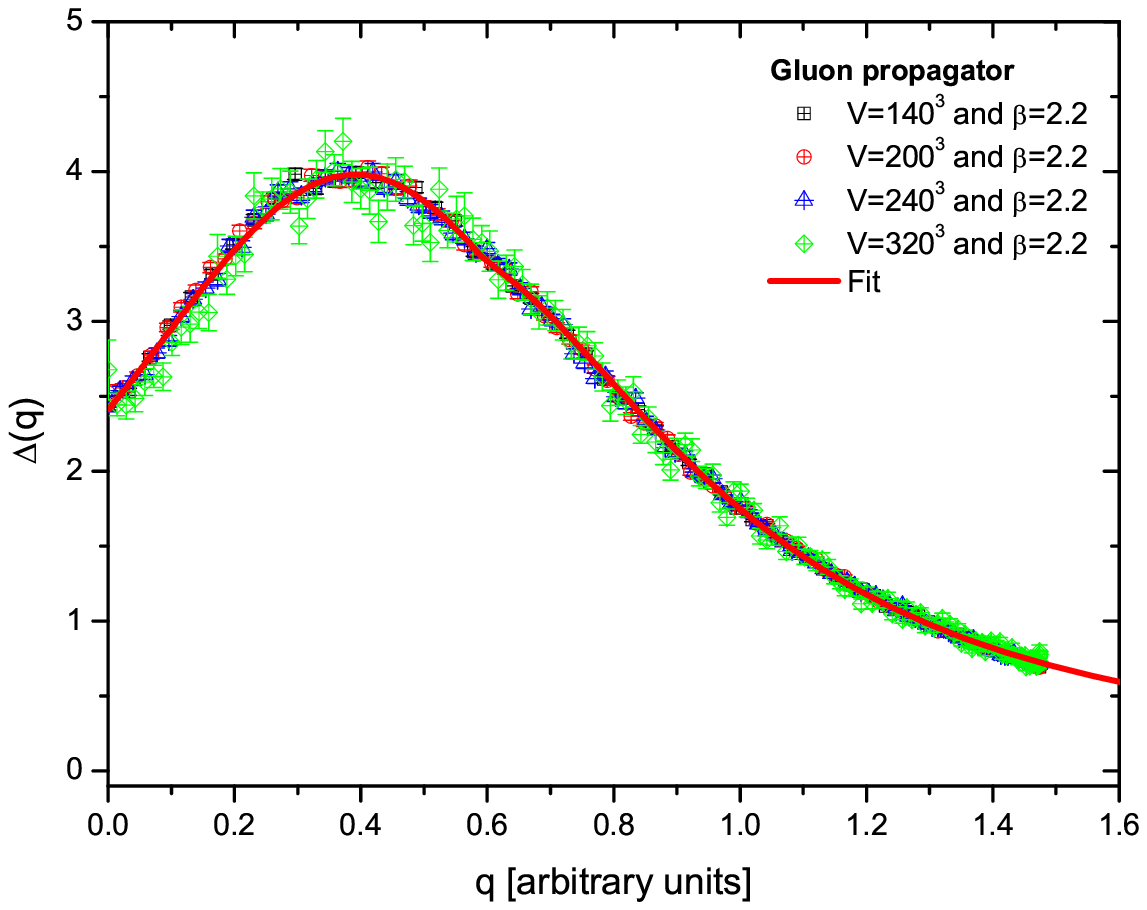}
\end{minipage}
\hspace{0.5cm}
\begin{minipage}[b]{0.50\linewidth}
%\hspace{-1.5cm}
\includegraphics[scale=0.5]{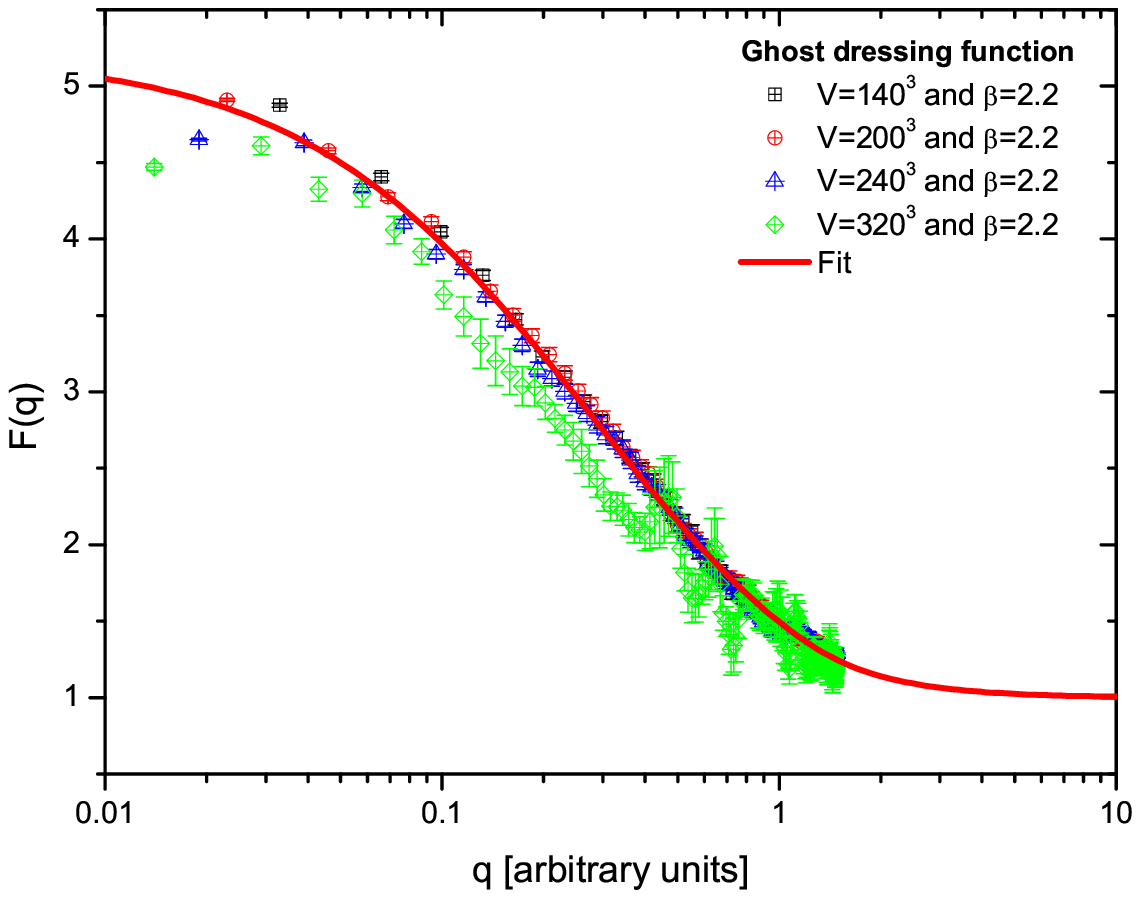}
\end{minipage}
\end{center}
\vspace{-0.5cm}
\caption{{\it Left panel}: Lattice results for the $SU(2)$ gluon 
propagator in $d=3$. The continuous 
line represents our best fit to the data
obtained from Eq.~(4.3).  {\it Right panel}: Lattice data for the $SU$(2) ghost 
dressing function $F(q)$ in 3 dimensions; the solid line 
corresponds to the best fit given by Eq.~(4.4).} 
\label{GdressandChi-3d}
\end{figure}

Now we will repeat the same exercise using the lattice
results for $d=3$ and $N=2$. Let us start, as in the previous case,  by showing in Fig.~\ref{GdressandChi-3d} 
the lattice results~\cite{Cucchieri:2003di}  for the three-dimensional gluon 
propagator $\Delta(q)$ (left panel) and the  ghost dressing function $F(q)$ (right panel). Both $\Delta(q)$ and $F(q)$
saturate in the deep  IR region, and can therefore be fitted by means of IR finite expressions. 

%%%%%%%%%%%%%%%%%%%%%%%%%%%%%%%%%%%%%%%%%%%%%%%%%%%%%%%%%%%%%%%%%%%%%%%%%%
%             Fig.4  gluon propagator and ghost dressing SU(2) in 3d
%%%%%%%%%%%%%%%%%%%%%%%%%%%%%%%%%%%%%%%%%%%%%%%%%%%%%%%%%%%%%%%%%%%%%%%%%%%%
%%%%%%%%%%%%%%%%%%%%%%%%%%%%%%%%%%%%%%%%%%%%%%%%%%%%%%%%%%%%%%%%%%%%

In the case of the gluon propagator, an accurate fit is giving by 
\be
\Delta(q) = A\exp\left[-(q-q_0)^2/w\right] + \frac{1}{a+bq + cq^2} \,, 
\label{fit3d}
\ee
where the fitting parameters 
are $A=0.49$, $q_0=0.11$, $w=0.37$, $a=0.43$, $b=-0.85$, and $c=1.143$. 
For the ghost dressing function, 
we use the following piecewise interpolator
\bea
F(q) = \frac1{a+bq+cq^2}, \,\,\, \mbox{for} \quad q^2 \leq 3 
\quad \mbox{and} \quad 
F(q)=  1 + \frac{d}{eq + q^2},  \,\,\, \mbox{for} \quad q^2 > 3 \,,
\label{ghdr3d-fit}
\eea
with fitting coefficients $a=0.19$, $b=0.61$, $c= -0.14$, $d=0.63$ and $e=0.26$ 
obtained by requiring the function to be continuous at $q^2=3$. 

\begin{figure}[!b]
\begin{center}
\begin{minipage}[b]{0.45\linewidth}
\centering
%\hspace{-1cm}
\includegraphics[scale=0.5]{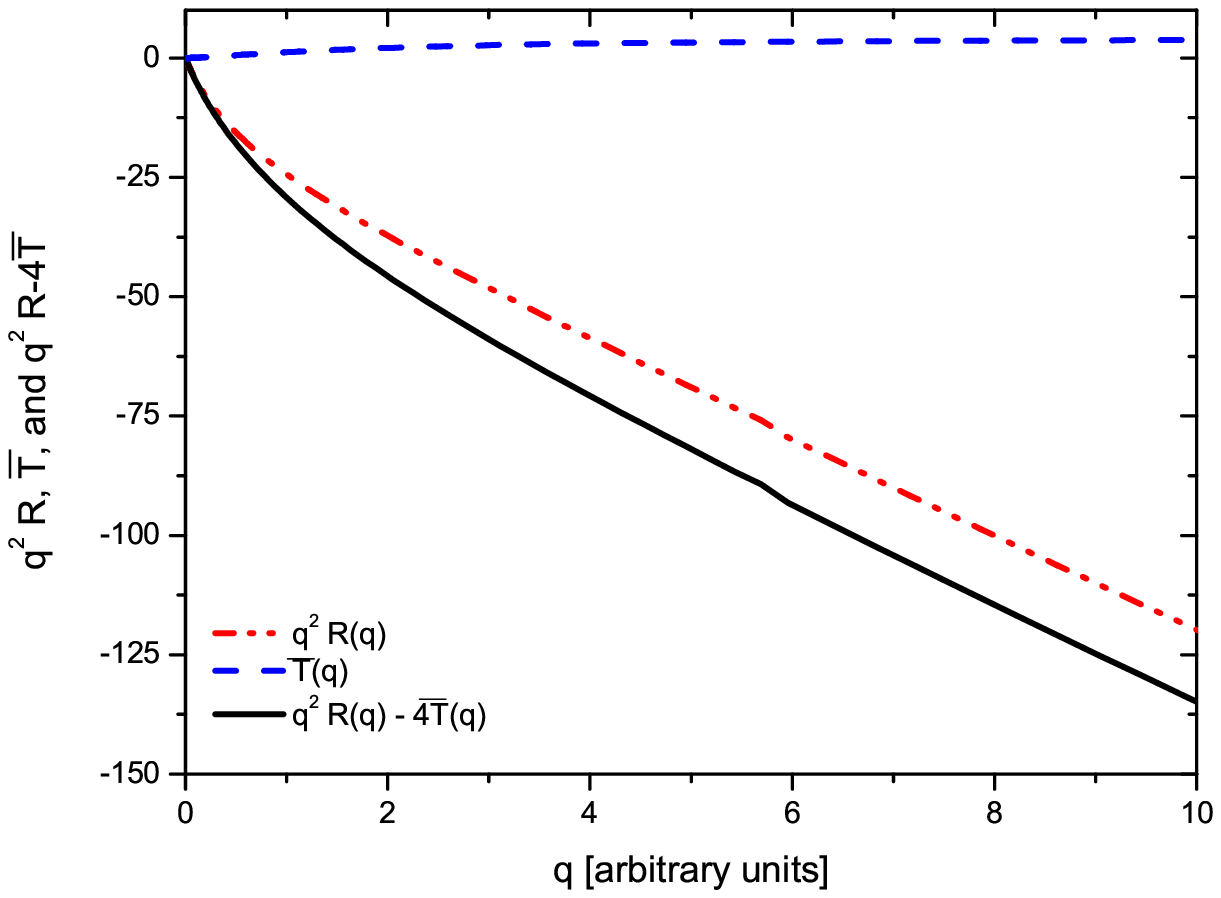}
\end{minipage}
\hspace{0.5cm}
\begin{minipage}[b]{0.50\linewidth}
%\hspace{-1.5cm}
\includegraphics[scale=0.5]{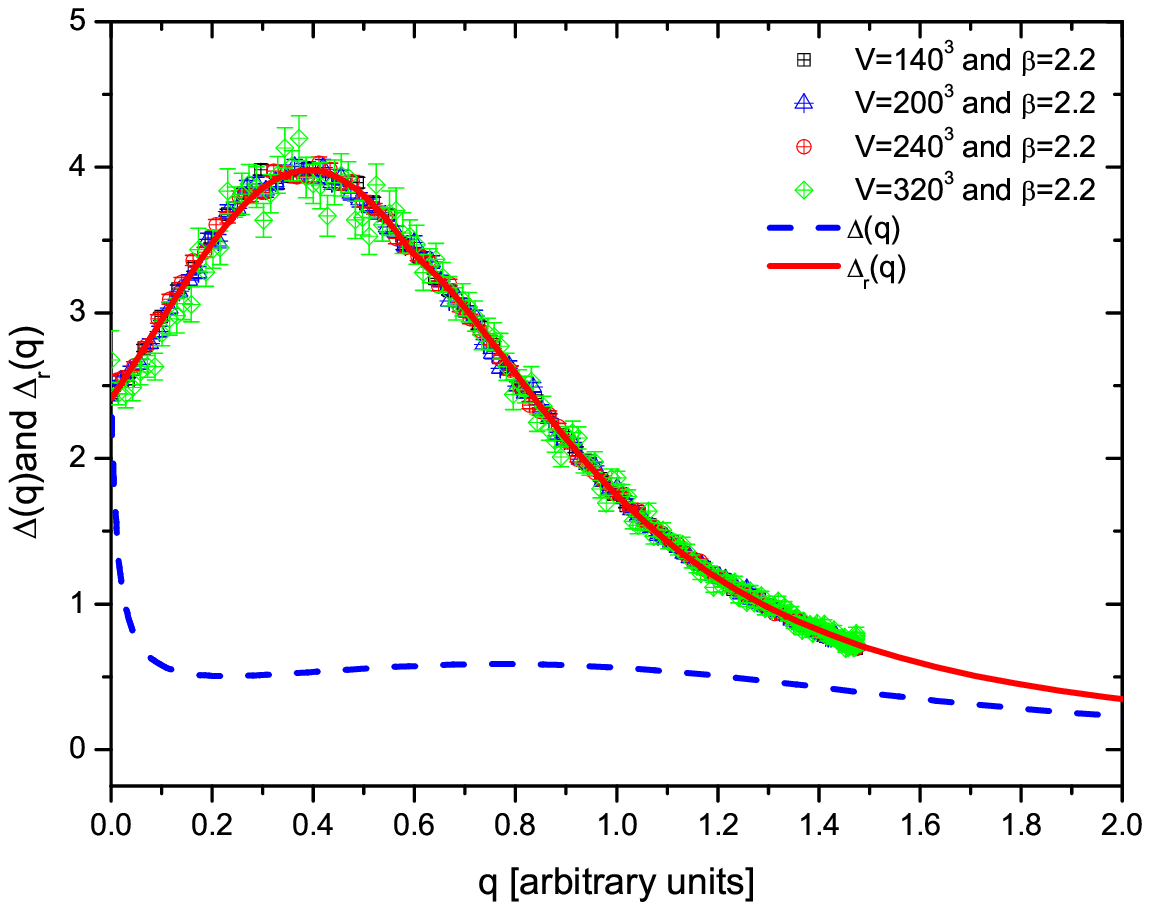}
\end{minipage}
\end{center}
\vspace{-0.5cm}
\caption{\label{RTandGluon-3d}{\it Left panel}: Numerical evaluation 
of the ghost contribution $\Pi_c$ to the gluon propagator using as 
input our best fit for the $d=3$, $N=2$ ghost dressing lattice 
data. {\it Right panel}: The result of removing the one-loop dressed 
ghost contribution from the gluon propagator in $d=3$. The effect is 
much more dramatic than in the $d=4$ case, since all the structure 
is determined by the ghost contribution, while $\Delta_r$ has the sole 
(but crucial!) role of rendering the propagator finite at $q=0$.} 
\end{figure}

Next, substituting the results presented on the left panel of  Fig.~\ref{RTandGluon-3d}
into Eqs.~(\ref{PicTR}) and (\ref{me}), and using  \mbox{$g=1.208$}, we compute $\Pi_c$ and $\Delta_r$.
On the right panel of Fig.~\ref{RTandGluon-3d}, 
we compare the residual propagator $\Delta_r$ (blue dashed line) with the full propagator $\Delta(q)$. Clearly, the 
effect in the tridimensional case 
is even more pronounced: the 
ghost contribution completely dominates over the rest, 
determining to a large extent the overall shape and structure of the propagator. 

%%%%%%%%%%%%%%%%%%%%%%%%%%%%%%%%%%%%%%%%%%%%%%%%%%%%%%%%%%%%%%%%%%%%%%%%%%
%             Fig.5  Pi_c and Delta_r SU(2) in 3d
%%%%%%%%%%%%%%%%%%%%%%%%%%%%%%%%%%%%%%%%%%%%%%%%%%%%%%%%%%%%%%%%%%%%%%%%%%%%
%%%%%%%%%%%%%%%%%%%%%%%%%%%%%%%%%%%%%%%%%%%%%%%%%%%%%%%%%%%%%%%%%%%%

\section{The effects of the ghost loop in the mass equation}

In the previous section we have studied how the subtraction 
of the ghost contributions affects the profile of the  gluon
propagator. However, as we will see now, the effects goes way  
beyond a simple change in the overall propagator shape, modifying its salient 
qualitative characteristics, and in particular the generation of a dynamical gluon mass.

To establish this, we use the $q\to 0$ limit of 
the equation describing the behavior of the dynamical mass
equation, {\it i.e.}  
\be
m^2(0)=-\frac{d-1}{d(4\pi)^{\frac d2}\Gamma
\left(\frac d2\right)}\frac{4g^2C_A}{1+G(0)}\int_0^\infty\!\diff y\,m^2(y){\cal K}_{d;N}(y),
\label{fnal}
\ee
with the kernel ${\cal K}_{d;N}$ given by
\be
{\cal K}_{d;N}(y)=y^{\frac d2-1}\Delta(y)[y\Delta(y)]'. 
\label{ker}
\ee

Since the constant multiplying the integral is positive, the 
negative sign in front of Eq.~(\ref{fnal}) tells us that the 
required physical constraint $m^2(0)>0$ can be fulfilled if and only 
if the integral kernel ${\cal K}_{d;N}$ (constructed solely out
of the gluon propagator) displays a sufficiently deep and extended negative
region at intermediate momenta~\cite{Aguilar:2011ux}.

%%%%%%%%%%%%%%%%%%%%%%%%%%%%%%%%%%%%%%%%%%%%%%%%%%%%%%%%%%%%%%%%%%%%%%%%%%
%             Fig.6 Kernels
%%%%%%%%%%%%%%%%%%%%%%%%%%%%%%%%%%%%%%%%%%%%%%%%%%%%%%%%%%%%%%%%%%%%%%%%%%%%
\begin{figure}[!t]
\begin{center}
\includegraphics[scale=0.9]{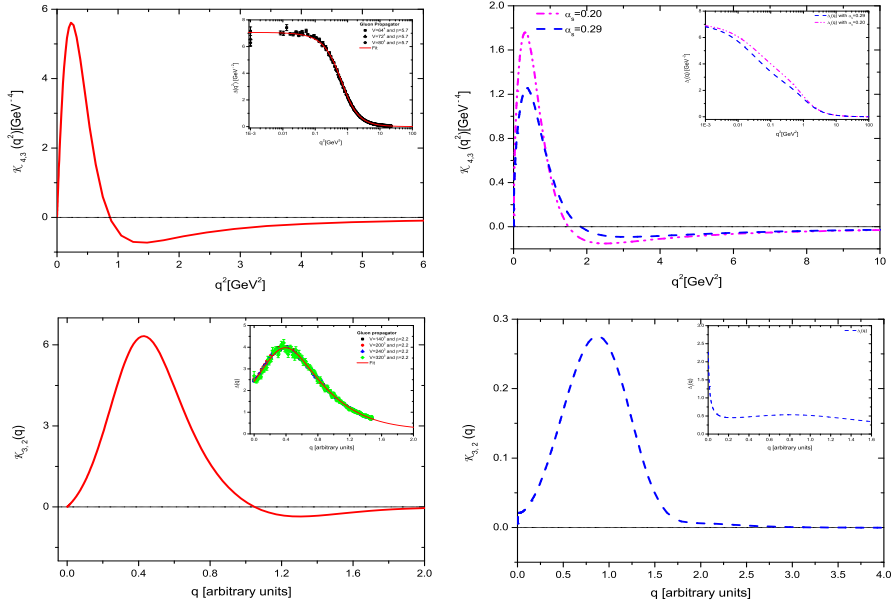}
\caption{\label{Kernel-all}The kernel ${\cal K}_{d;N}$ of Eq.~(5.2) 
constructed out of the lattice propagator $\Delta$ (left panels) and
the ghost-less propagator $\Delta_r$ (right panels) for the $d=4$ $N=3$ 
(top row), and  $d=3$ $N=2$ (bottom row) cases. The insets
show in each case the shape of the propagator used to evaluate the kernels.}
\end{center}
\end{figure}
%%%%%%%%%%%%%%%%%%%%%%%%%%%%%%%%%%%%%%%%%%%%%%%%%%%%%%%%%%%%%%%%%%%%

In the left panels of Fig.~\ref{Kernel-all} we plot the 
kernels ${\cal K}_{d;N}$ obtained from the lattice data for the
cases  $d=4$, $N=3$  (top row), and $d=3$ $N=2$ (bottom row); both 
 display the characteristic negative region that allows, at least
in principle, the existence of solutions of Eq.~(\ref{fnal}), furnishing 
a positive value for $m^2(0)$.

On the other hand, the situation changes substantially once 
the ghost loop is removed, in which case the kernels ${\cal K}_{d;N}$ must be 
constructed from $\Delta_r$ (right panels of the same figure). For $d=4$ one 
observes a shift towards higher $q$s of the zero crossing, and a correspondingly
suppressed negative region; even though this is not sufficient to
exclude {\it per se} the existence of a physical solution to the mass
equation~(\ref{fnal}), a thorough study of the approximate
equation derived in~\cite{Aguilar:2011ux} reveals that no physical solution 
may be found. The $d=3$ situation is even more obvious: the highly suppressed negative region present 
in this case cannot support solutions of~(\ref{fnal}) with $m^2(0)>0$, thus
leaving as the only possibility the trivial $m^2=0$ solution.

The main conclusion one can draw, therefore, is that the ghosts play 
a fundamental role in the mechanism of 
dynamical gluon mass generation, since the failure to
properly include them results in the inability of the theory to generate
dynamically a mass for the gluon.

\section{Conclusions}

In this talk  we have presented 
a study of the impact of the ghost sector  
on the overall form of the gluon propagator in a pure Yang-Mills theory,   
for different space-time dimensions ($d=3,\ 4$) and $SU(N)$ gauge groups ($N=2,3$).

The suppression of the gluon propagator induced by the removal of the ghost-loops 
has far-reaching consequences on the mechanism that endows gluons with a 
dynamical mass, associated with the  observed IR-finiteness of 
the gluon propagator and the ghost-dressing function.
Specifically, using a recently 
derived integral equation controlling the dynamics of the (momentum-dependent) gluon mass,
we have demonstrated that when the  reduced gluon propagators are used as inputs, 
the corresponding kernels are  modified in such a way that no 
physical solutions may be found, thus failing to generate a mass gap for 
the pure Yang-Mills theory~\cite{Aguilar:2011yb}. Instead, as has been shown in~\cite{Aguilar:2011ux}, the 
use of the full gluon propagator in the same equation generates a physically acceptable 
gluon mass.

\acknowledgments
I would like to thank the ECT* for the hospitality and for supporting
the QCD-TNT II organization. This work was supported by Funda\c{c}\~ao de Amparo \`a Pesquisa do Estado de S\~ao Paulo (Fapesp)
under grant 2011/11474-8 and by the Brazilian Funding Agency CNPq - grant 305850/2009-1.


\begin{thebibliography}{99}



%\cite{Cucchieri:2007md}
\bibitem{Cucchieri:2007md}
A.~Cucchieri and T.~Mendes,
%``What's up with IR gluon and ghost propagators in Landau gauge? A puzzling
%answer from huge lattices,''
PoS {\bf LAT2007}, 297 (2007);
%  [arXiv:0710.0412 [hep-lat]].
%%CITATION = POSCI,LAT2007,297;%%
%
%\cite{Cucchieri:2007rg}
%\bibitem{Cucchieri:2007rg}
%A.~Cucchieri and T.~Mendes,
%``Constraints on the IR behavior of the gluon propagator in Yang-Mills
%theories,''
Phys.\ Rev.\ Lett.\  {\bf 100}, 241601 (2008).
%  [arXiv:0712.3517 [hep-lat]].
%%CITATION = PRLTA,100,241601;%%  


%\cite{Cucchieri:2003di}
\bibitem{Cucchieri:2003di}
  A.~Cucchieri, T.~Mendes and A.~R.~Taurines,
  %``SU(2) Landau gluon propagator on a 140^3 lattice,''
  Phys.\ Rev.\  D {\bf 67}, 091502 (2003).
  %[arXiv:hep-lat/0302022].
  %%CITATION = PHRVA,D67,091502;%%


%\cite{Bogolubsky:2007ud}
\bibitem{Bogolubsky:2007ud}
 I.~L.~Bogolubsky, E.~M.~Ilgenfritz, M.~Muller-Preussker and A.~Sternbeck,
  %``The Landau gauge gluon and ghost propagators in 4D SU(3) gluodynamics in
%large lattice volumes,''
PoS {LATTICE}, 290 (2007).
%  [arXiv:0710.1968 [hep-lat]].
%%CITATION = POSCI,LATTICE,290;%%


%\cite{Oliveira:2008uf}
\bibitem{Oliveira:2008uf}
  O.~Oliveira, P.~J.~Silva,
  %``Does The Lattice Zero Momentum Gluon Propagator for Pure Gauge SU(3) Yang-Mills Theory Vanish in the Infinite Volume Limit?,''
  Phys.\ Rev.\  {\bf D79}, 031501 (2009);
%  [arXiv:0809.0258 [hep-lat]].
%\cite{Oliveira:2009eh}
%\bibitem{Oliveira:2009eh}
%O.~Oliveira and P.~J.~Silva,
%``The lattice infrared Landau gauge gluon propagator: the infinite volume
%limit,''
PoS {\bf LAT2009}, 226 (2009).
% [arXiv:0910.2897 [hep-lat]].
%%CITATION = POSCI,LAT2009,226;%%


%\cite{arXiv:0912.0437}
\bibitem{arXiv:0912.0437} 
  H.~Suganuma, T.~Iritani, A.~Yamamoto and H.~Iida,
  %``Lattice QCD Analysis for Gluons,''
  PoSQCD\ {\bf -TNT09}, 044  (2009); 
%  [arXiv:0912.0437 [hep-lat]].
%\cite{arXiv:1011.0007}
%\bibitem{arXiv:1011.0007} 
%  H.~Suganuma, T.~Iritani, A.~Yamamoto and H.~Iida,
  %``Lattice QCD Study for Gluon Propagator and Gluon Spectral Function,''
  PoSLATTICE\ {\bf 2010}, 289  (2010).
 % [arXiv:1011.0007 [hep-lat]].
  %%CITATION = POSCI,LATTICE2010,289;%%
  
%\cite{Bowman:2007du}
\bibitem{Bowman:2007du}
P.~O.~Bowman {\it et al.},
%``Scaling behavior and positivity violation of the gluon propagator 
%in full QCD,''
Phys.\ Rev.\  D {\bf 76}, 094505 (2007).
%  [arXiv:hep-lat/0703022].
%%CITATION = PHRVA,D76,094505;%%
  

%%%%%%%%
%  SDE
%%%%%%%%

%\cite{Aguilar:2008xm}
\bibitem{Aguilar:2008xm}
  A.~C.~Aguilar, D.~Binosi and J.~Papavassiliou,
  %``Gluon and ghost propagators in the Landau gauge: Deriving lattice results
  %from Schwinger-Dyson equations,''
  Phys.\ Rev.\  D {\bf 78}, 025010 (2008);
  %[arXiv:0802.1870 [hep-ph]].
  %%CITATION = PHRVA,D78,025010;%%
 %\cite{Aguilar:2010zx}
%\bibitem{Aguilar:2010zx}
  A.~C.~Aguilar, D.~Binosi and J.~Papavassiliou,
  %``Nonperturbative gluon and ghost propagators for d=3 Yang-Mills,''
  Phys.\ Rev.\  D {\bf 81}, 125025 (2010).
  %[arXiv:1004.2011 [hep-ph]].
  %%CITATION = PHRVA,D81,125025;%%

%\cite{Binosi:2009qm}
\bibitem{Binosi:2009qm}  
D.~Binosi and J.~Papavassiliou,
%``Pinch Technique: Theory and Applications,''
Phys.\ Rept.\  {\bf 479}, 1-152 (2009).
%[arXiv:0909.2536 [hep-ph]].

  
%\cite{Boucaud:2008ji}
\bibitem{Boucaud:2008ji}
Ph.~Boucaud, J.~P.~Leroy, A.~L.~Yaouanc, J.~Micheli, O.~Pene and J.~Rodriguez-Quintero,
%``IR finiteness of the ghost dressing function from numerical resolution of
%the ghost SD equation,''
JHEP {\bf 0806} (2008) 012.
%[arXiv:0801.2721 [hep-ph]];
%%CITATION = JHEPA,0806,012;%%  

%%%%%
% functional methods
%%%%%%
%\cite{Braun:2007bx}
\bibitem{Braun:2007bx}
J.~Braun, H.~Gies and J.~M.~Pawlowski,
%``Quark Confinement from Color Confinement,''
Phys.\ Lett.\  B {\bf 684}, 262 (2010).
%  [arXiv:0708.2413 [hep-th]].
%%CITATION = PHLTA,B684,262;%%

%\cite{Szczepaniak:2010fe}
\bibitem{Szczepaniak:2010fe}
A.~P.~Szczepaniak and H.~H.~Matevosyan,
%``A model for QCD ground state with magnetic disorder,''
Phys.\ Rev.\  D {\bf 81}, 094007 (2010).
%  [arXiv:1003.1901 [hep-ph]].
%%CITATION = PHRVA,D81,094007;%%


%\cite{Zwanziger:1993dh}
\bibitem{Zwanziger:1993dh}
  D.~Zwanziger,
  %``Fundamental modular region, Boltzmann factor and area law in lattice gauge
  %theory,''
  Nucl.\ Phys.\  B {\bf 412}, 657 (1994).
  %%CITATION = NUPHA,B412,657;%%
%%%%%%%%%

%\cite{Dudal:2008sp}
\bibitem{Dudal:2008sp}
D.~Dudal, J.~A.~Gracey, S.~P.~Sorella, N.~Vandersickel and H.~Verschelde,
%``A refinement of the Gribov-Zwanziger approach in the Landau gauge: infrared
%propagators in harmony with the lattice results,''
Phys.\ Rev.\  D {\bf 78}, 065047 (2008).
%  [arXiv:0806.4348 [hep-th]].
%%CITATION = PHRVA,D78,065047;%%


%\cite{Kondo:2011ab}
\bibitem{Kondo:2011ab} 
  K.~-I.~Kondo,
  %``A low-energy effective Yang-Mills theory for quark and gluon confinement,''
  Phys.\ Rev.\ D {\bf 84}, 061702 (2011).
 % [arXiv:1103.3829 [hep-th]].
  %%CITATION = ARXIV:1103.3829;%%
  
%%%%%%%%
%  gluon mass
%%%%%%%%

%\cite{Cornwall:1981zr}
\bibitem{Cornwall:1981zr}
J.~M.~Cornwall,
%``Dynamical Mass Generation In Continuum QCD,''
Phys.\ Rev.\ D {\bf 26}, 1453 (1982).
%%CITATION = PHRVA,D26,1453;%%


%%%%%%
%  ghost dominance
%%%%%%

%\cite{Alkofer:2000wg}
\bibitem{Alkofer:2000wg}
 R.~Alkofer, L.~von Smekal,
 %``The Infrared behavior of QCD Green's functions: Confinement 
% dynamical symmetry breaking, and hadrons as relativistic bound states,''
 Phys.\ Rept.\  {\bf 353}, 281 (2001).
%  [hep-ph/0007355].


%\cite{Zwanziger:2001kw}
\bibitem{Zwanziger:2001kw}
  D.~Zwanziger,
  %``Nonperturbative Landau gauge and infrared critical exponents in QCD,''
  Phys.\ Rev.\  {\bf D65}, 094039 (2002).
  %[hep-th/0109224].


%\cite{Fischer:2003rp}
\bibitem{Fischer:2003rp}
  C.~S.~Fischer, R.~Alkofer,
  %``Nonperturbative propagators, running coupling and dynamical quark mass of Landau gauge QCD,''
  Phys.\ Rev.\  {\bf D67}, 094020 (2003).
%  [hep-ph/0301094].


%\cite{Aguilar:2010cn}
\bibitem{Aguilar:2010cn}
A.~C.~Aguilar and J.~Papavassiliou,
%``Chiral symmetry breaking with lattice propagators,''
Phys.\ Rev.\ D {\bf 83}, 014013 (2011).
%  [arXiv:1010.5815 [hep-ph]].
%%CITATION = PHRVA,D83,014013;%%


%\cite{Aguilar:2011yb}
\bibitem{Aguilar:2011yb} 
  A.~C.~Aguilar, D.~Binosi and J.~Papavassiliou,
  %``Gluon mass through ghost synergy,''
  arXiv:1108.5989 [hep-ph].
  %%CITATION = ARXIV:1108.5989;%%


%\cite{Cornwall:1989gv}
\bibitem{Cornwall:1989gv}
J.~M.~Cornwall and J.~Papavassiliou,
%``Gauge Invariant Three Gluon Vertex in QCD,''
Phys.\ Rev.\  D {\bf 40}, 3474 (1989).
%%CITATION = PHRVA,D40,3474;%%


%\cite{Aguilar:2006gr}
\bibitem{Aguilar:2006gr}
A.~C.~Aguilar and J.~Papavassiliou,
%``Gluon mass generation in the PT-BFM scheme,''
JHEP {\bf 0612}, 012 (2006).
%  [arXiv:hep-ph/0610040].
%%CITATION = JHEPA,0612,012;%%

%\cite{Binosi:2007pi}
\bibitem{Binosi:2007pi}
D.~Binosi and J.~Papavassiliou,
%``Gauge-invariant truncation scheme for the Schwinger-Dyson equations of
%QCD,''
Phys.\ Rev.\  D {\bf 77}(R), 061702 (2008);
%%CITATION = PHRVA,D77,061702;%% %\cite{Binosi:2008qk}
%\bibitem{Binosi:2008qk}
%D.~Binosi and J.~Papavassiliou,
%``New Schwinger-Dyson equations for non-Abelian gauge theories,''
JHEP {\bf 0811}, 063 (2008).
%[arXiv:0805.3994 [hep-ph]].
%%CITATION = JHEPA,0811,063;%%


%\cite{Abbott:1980hw}
\bibitem{Abbott:1980hw}
See, e.g.,  L.~F.~Abbott,
%``The Background Field Method Beyond One Loop,''
Nucl.\ Phys.\  B {\bf 185}, 189 (1981), and references therein.
%%CITATION = NUPHA,B185,189;%%

%\cite{Aguilar:2009nf}
\bibitem{Aguilar:2009nf}
  A.~C.~Aguilar, D.~Binosi, J.~Papavassiliou and J.~Rodriguez-Quintero,
  %``Non-perturbative comparison of QCD effective charges,''
  Phys.\ Rev.\  D {\bf 80}, 085018 (2009).
%  [arXiv:0906.2633 [hep-ph]].
  %%CITATION = PHRVA,D80,085018;%%


%\cite{Grassi:2004yq}
\bibitem{Grassi:2004yq}
  P.~A.~Grassi, T.~Hurth and A.~Quadri,
  %``On the Landau background gauge fixing and the IR properties of YM Green
  %functions,''
  Phys.\ Rev.\  D {\bf 70}, 105014 (2004).
%  [arXiv:hep-th/0405104].
  %%CITATION = PHRVA,D70,105014;%%


%\cite{Aguilar:2009pp}
\bibitem{Aguilar:2009pp}
  A.~C.~Aguilar, D.~Binosi and J.~Papavassiliou,
  %``Indirect determination of the Kugo-Ojima function from lattice data,''
  JHEP {\bf 0911}, 066 (2009).
 % [arXiv:0907.0153 [hep-ph]].
  %%CITATION = JHEPA,0911,066;%%


%\cite{Salam:1963sa}
\bibitem{Salam:1963sa}
  A.~Salam,
  %``Renormalizable electrodynamics of vector mesons,''
  Phys.\ Rev.\  {\bf 130}, 1287 (1963);
  %%CITATION = PHRVA,130,1287;%%
  %\cite{Salam:1964zk}
%\bibitem{Salam:1964zk}
  A.~Salam and R.~Delbourgo,
  %``Renormalizable electrodynamics of scalar and vector mesons. II,''
  Phys.\ Rev.\  {\bf 135}, B1398 (1964);
  %%CITATION = PHRVA,135,B1398;%%
%\cite{Delbourgo:1977jc}
%\bibitem{Delbourgo:1977jc}
  R.~Delbourgo and P.~C.~West,
  %``A Gauge Covariant Approximation To Quantum Electrodynamics,''
  J.\ Phys.\ A  {\bf 10}, 1049 (1977);
  %%CITATION = JPAGB,A10,1049;%%
%\cite{Delbourgo:1977hq}
%\bibitem{Delbourgo:1977hq}
  R.~Delbourgo and P.~C.~West,
  %``Infrared Behavior Of A Gauge Covariant Approximation,''
  Phys.\ Lett.\  B {\bf 72}, 96 (1977).
  %%CITATION = PHLTA,B72,96;%%

%\cite{Ball:1980ay}
\bibitem{Ball:1980ay}
  J.~S.~Ball, T.~-W.~Chiu,
  %``Analytic Properties of the Vertex Function in Gauge Theories. 1.,''
  Phys.\ Rev.\  {\bf D22}, 2542 (1980).

%\cite{Kizilersu:2009kg}
\bibitem{Kizilersu:2009kg}
  A.~Kizilersu and M.~R.~Pennington,
  %``Building the Full Fermion-Photon Vertex of QED by Imposing Multiplicative
  %Renormalizability of the Schwinger-Dyson Equations for the Fermion and Photon
  %Propagators,''
  Phys.\ Rev.\  D {\bf 79}, 125020 (2009);
 % [arXiv:0904.3483 [hep-th]].
  %%CITATION = PHRVA,D79,125020;%%
%\cite{Bashir:1997qt}
%\bibitem{Bashir:1997qt}
  A.~Bashir, A.~Kizilersu and M.~R.~Pennington,
  %``The non-perturbative three-point vertex in massless quenched QED and
  %perturbation theory constraints,''
  Phys.\ Rev.\  D {\bf 57}, 1242 (1998).
%  [arXiv:hep-ph/9707421].
  %%CITATION = PHRVA,D57,1242;%%  


%\cite{Aguilar:2009ke}
\bibitem{Aguilar:2009ke}
  A.~C.~Aguilar and J.~Papavassiliou,
  %``Gluon mass generation without seagull divergences,''
  Phys.\ Rev.\  D {\bf 81}, 034003 (2010).
 % [arXiv:0910.4142 [hep-ph]].
  %%CITATION = PHRVA,D81,034003;%%

%\cite{Aguilar:2011ux}
\bibitem{Aguilar:2011ux} 
  A.~C.~Aguilar, D.~Binosi and J.~Papavassiliou,
  %``The dynamical equation of the effective gluon mass,''
  Phys.\ Rev.\ D {\bf 84}, 085026 (2011).
%  [arXiv:1107.3968 [hep-ph]].
  %%CITATION = ARXIV:1107.3968;%%


%\cite{Aguilar:2010gm}
\bibitem{Aguilar:2010gm}
  A.~C.~Aguilar, D.~Binosi, J.~Papavassiliou,
  %``QCD effective charges from lattice data,''
  JHEP {\bf 1007}, 002 (2010).
 % [arXiv:1004.1105 [hep-ph]].


%\cite{Lavelle:1991ve}
\bibitem{Lavelle:1991ve}
  M.~Lavelle,
  %``Gauge invariant effective gluon mass from the operator product expansion,''
  Phys.\ Rev.\  {\bf D44}, 26-28 (1991).

%\cite{Aguilar:2007ie}
\bibitem{Aguilar:2007ie}
  A.~C.~Aguilar, J.~Papavassiliou,
  %``Power-law running of the effective gluon mass,''
  Eur.\ Phys.\ J.\  {\bf A35}, 189-205 (2008).
%  [arXiv:0708.4320 [hep-ph]].

%\cite{Oliveira:2010xc}
\bibitem{Oliveira:2010xc}
  O.~Oliveira, P.~Bicudo,
  %``Running Gluon Mass from Landau Gauge Lattice QCD Propagator,''
  J.\ Phys.\ G {\bf G38}, 045003 (2011).
%  [arXiv:1002.4151 [hep-lat]].

  
  


\end{thebibliography}
\end{document}